# An absorption profile centred at 78 MHz in the sky-averaged spectrum

Judd D. Bowman[1], Alan E. E. Rogers[2], Raul A. Monsalve[3,1,4], Thomas J. Mozdzen[1], Nivedita Mahesh[1]

[1]School of Earth and Space Exploration, Arizona State University, Tempe, Arizona 85287, USA
[2] Haystack Observatory, Massachusetts Institute of Technology, Westford, Massachusetts 01886, USA
[3]Center for Astrophysics and Space Astronomy, University of Colorado, Boulder, Colorado 80309, USA
[4]Facultad de Ingeniería, Universidad Católica de la Santísima Concepción, Alonso de Ribera 2850, Concepción, Chile

**After stars formed in the early Universe, their ultraviolet light is expected, eventually, to have penetrated the primordial hydrogen gas and altered the excitation state of its 21-centimetre hyperfine line. This alteration would cause the gas to absorb photons from the cosmic microwave background, producing a spectral distortion that should be observable today at radio frequencies of less than 200 megahertz[1]. Here we report the detection of a flattened absorption profile in the sky-averaged radio spectrum, which is centred at a frequency of 78 megahertz and has a best-fitting full-width at half-maximum of 19 megahertz and an amplitude of 0.5 kelvin. The profile is largely consistent with expectations for the 21-centimetre signal induced by early stars; however, the best-fitting amplitude of the profile is more than a factor of two greater than the largest predictions[2]. This discrepancy suggests that either the primordial gas was much colder than expected or the background radiation temperature was hotter than expected. Astrophysical phenomena (such as radiation from stars and stellar remnants) are unlikely to account for this discrepancy; of the proposed extensions to the standard model of cosmology and particle physics, only cooling of the gas as a result of interactions between dark matter and baryons seems to explain the observed amplitude[3]. The low-frequency edge of the observed profile indicates that stars existed and had produced a background of Lyman-α photons by 180 million years after the Big Bang. The high-frequency edge indicates that the gas was heated to above the radiation temperature less than 100 million years later.**

Observations with the Experiment to Detect the Global EoR Signature (EDGES) low-band instruments beginning in August 2015 were used to detect the absorption profile. Each of the two low-band instruments consists of a radio receiver and a zenith-pointing, single-polarisation dipole antenna. Spectra of the radio sky-noise brightness temperature, spatially averaged over the large beams of the instruments, were recorded between 50 and 100 MHz. Raw spectra were calibrated, filtered, and integrated over hundreds of hours. Automated antenna reflection coefficient (S11) measurement was performed in the field. Low-noise amplifier (LNA) noise waves and S11 were measured in the laboratory, along with additional calibration constants. Details of the instruments, calibration, verification, and model fitting are described in the Methods section.

In Figure 1 we summarize the detection. It shows the spectrum observed by one of the instruments and the results of model fits. Galactic synchrotron emission dominates the observed sky noise yielding a power-law spectral profile that falls from ~5000 K at 50 MHz to ~1000 K at 100 MHz for the high Galactic latitudes shown. Fitting and removing the Galactic emission and ionospheric contributions from the spectrum using a five-term, physically motivated foreground model (equation (1) in Methods) results in a residual with root-mean-square (r.m.s.) of 0.087 K. The absorption profile is found by fitting the integrated spectrum with the foreground model and a model for the 21-cm signal simultaneously. The best-fitting 21-cm model yields a symmetric U-shaped absorption profile that is centred at a frequency of 78±1 MHz and has a full-width at half-maximum of $19^{+4}_{-2}$ MHz, an amplitude of $0.5^{+0.5}_{-0.2}$ K and a flattening factor of $\tau = 7^{+5}_{-3}$ (where the bounds provide 99% confidence intervals including estimates of systematic uncertainties; see Methods for model definition). Uncertainties in the parameters of the fitted profile are estimated from statistical uncertainty in the model fits and from systematic differences between the various validation trials that were performed using observations from both instruments and several different data cuts. The 99% confidence intervals that we report are calculated as the outer bounds of (1) the marginalized statistical 99% confidence intervals from fits to the primary dataset and (2) the range of best-fitting values for each parameter across the validation trials. Fitting with both the foreground and 21-cm models lowers the residuals to an r.m.s. of 0.025 K. The fit shown in Figure 1 has a signal-to-noise ratio of 37, calculated as the best-fitting amplitude of the profile divided by the statistical uncertainty of the amplitude fit, including the covariance between model parameters. Additional analyses of the observations using restricted spectral bands yield nearly identical best-fit absorption profiles with the highest SNR reaching 52. In Figure 2 we show representative cases of these fits.

We performed numerous hardware and processing tests to validate the detection. The 21 cm absorption profile is observed in data spanning nearly two years and can be extracted at all local solar times and at all local sidereal times (LST). The absorption profile is detected by the two identically-designed instruments operated at the same site and located 150 meters apart. It is detected with several hardware modifications to the instruments, including orthogonal orientations of one of the antennas. Similar results for the absorption profile are obtained with two independent processing pipelines. The pipelines have been tested with simulated data. The profile is detected using data processed with two different calibration techniques, using calibration solutions from several laboratory measurements of the receivers, and using multiple on-site measurements of the antenna reflection coefficients. Sensitivity of the detection to several possible calibration errors has been modeled and in all cases recovered profile amplitudes are within the reported confidence range, as summarized in Table 1. An EDGES high-band instrument operates between 90 and 200 MHz at the same site using a nearly identical receiver and a scaled-version of the low-band antennas. It does not produce a similar feature at the scaled frequencies[4]. Analysis of radio-frequency interference (RFI) in the observations, including in the FM radio band, shows the absorption profile is inconsistent with typical spectral contributions from these sources.

We are not aware of alternative astronomical or atmospheric mechanisms capable of producing the observed profile. HII regions in the Galaxy have increasing optical depth with wavelength, blocking more background emission at lower frequencies, but they are observed primarily along the Galactic plane and generate monotonic spectral profiles at the observed frequencies. Radio recombination lines in the Galactic plane create a picket fence of narrow absorption lines separated by approximately 0.5 MHz at the observed frequencies [5], but they are easy to identify and filter in EDGES observations. The Earth's ionosphere weakly absorbs radio signals at the observed frequencies and emits thermal radiation from hot electrons, but models and observations show a broadband effect that varies depending on ionospheric conditions [6, 7], including diurnal changes in total electron content. This effect is fit by our foreground model. Molecules of the hydroxl radical and nitric oxide have spectral lines in the observed band and are present in the atmosphere, but the densities and line strengths are too low to produce significant absorption.

The 21 cm line has a rest-frame frequency of 1420 MHz. Expansion of the universe redshifts the line to the observed band according to $\nu = 1420 / (1+z)$ MHz, where z is redshift and maps uniquely to age of the Universe. The observed absorption profile is the continuous superposition of lines from gas across the observed redshift range and cosmological volume, hence the shape of the profile traces the history of the gas across cosmic time and is not the result of the properties of an individual cloud. The observed absorption profile is centred at $z \approx 17$ and spans approximately $20 > z > 15$.

The intensity of the observable 21 cm signal from the early Universe is given as a differential brightness temperature relative to the microwave background according to [8]:

$$T_{21}(z) \approx 0.023 \, x_{HI}(z) \left[ \left( \frac{0.15}{\Omega_M} \right) \left( \frac{1+z}{10} \right) \right]^{\frac{1}{2}} \left( \frac{\Omega_b h}{0.02} \right) \left[ 1 - \frac{T_R(z)}{T_S(z)} \right] \ \text{K}, \qquad (1)$$

where $x_{HI}$ is the hydrogen neutral fraction, $\Omega_M$ and $\Omega_b$ are the matter and baryon density relative to the critical density, respectively, $h$ is the Hubble constant relative to 100 km/s/Mpc, $T_R$ is the background radiation temperature, usually assumed to be from the background produced by the afterglow of the Big Bang, $T_S$ is the 21 cm spin temperature that defines the relative population of the hyperfine energy levels, and the factor of 0.023 K comes from atomic line physics and average gas density. The spin temperature is affected by absorption of microwave photons, which couples $T_S$ to $T_R$, as well as by resonant scattering of Lyman-$\alpha$ photons and atomic collisions, both of which couple $T_S$ to the gas kinetic temperature, $T_G$.

Gas and background radiation temperatures are coupled in the early Universe through Compton scattering. This coupling becomes ineffective in numerical models [9, 10] at $z \sim 150$, after which primordial gas cools adiabatically. In the absence of stars or non-standard physics, the gas temperature is expected to be 9.3 K at z=20 falling to 5.4 K at z=15. The radiation temperature cools more slowly due to cosmological expansion following $T_0(1+z)$, with $T_0 = 2.725$, and reaches 57.2 K and 43.6 K at the same redshifts, respectively.

Over time Lyman-$\alpha$ photons from early stars recouple the spin and gas temperatures [11], leading to the detected signal. The z=20 onset of the observed absorption profile places this epoch at an age of 180 million years, using the Planck 2015 cosmological parameters [12] . For the most extreme case, in which $T_S$ is fully coupled to $T_G$, the standard model yields a maximum absorption amplitude of 0.20 K at z=20 increasing to 0.23 K at z=15.

The presence of stars should eventually halt the cooling of the gas and ultimately heat it because stellar radiation deposits energy into the gas and Lyman line cooling has been modeled to be very small for expected stellar properties [13] . As early stars die, they are expected to leave behind stellar remnants, such as black holes and neutron stars. The accretion disks around these remnants should generate X-rays, further heating the gas. At some point, the gas is expected to become hotter than the background radiation temperature, ending the absorption signal. The z=15 edge of the observed profile places this transition around 270 million years after the Big Bang.

The ages derived for the events above fall within the range expected in many theoretical models [2] . The flattened shape of the observed absorption profile is uncommon in existing models, however, and could indicate that the initial flux of Lyman-$\alpha$ radiation from early stars was sufficiently large to quickly saturate the spin temperature to the gas temperature. High Lyman-$\alpha$ flux models were probed at z<14 using EDGES high-band measurements and a large fraction were found to be inconsistent with the data [4] .

In order to produce the best-fit profile amplitude of 0.5 K, the ratio of $T_R/T_S$ at the centre of the profile must be larger than 15, compared to 7 in the standard scenario. Even the lower confidence bound of 0.3 K for the observed profile amplitude is ~50% larger than the strongest predicted signal. For a standard gas temperature history, $T_R$ would need to be larger than 104 K to yield the best-fit amplitude at the centre of the profile, whereas for a radiation temperature history given solely by the microwave background, $T_G$ would need to be less than 3.2 K.

The observed profile amplitude could be explained if gas and background radiation temperatures decouple by z~250 rather than z~150, allowing the gas to begin cooling adiabatically earlier. A residual ionization fraction after the formation of atoms that is lower than expected by nearly an order of magnitude would lead to sufficiently early decoupling. However, cross-validation between numerical models and their consistency with Planck observations suggests that the residual ionization fraction is already known to ~1% fractional accuracy.

Considering more exotic scenarios, dark matter-baryon interactions can explain the observed profile amplitude by lowering the gas temperature [14] if the dark matter particle mass is below a few GeV and the interaction cross-section is greater than ~$10^{-21}$ cm$^2$, as derived in the companion paper [3] . Existing models of other non-standard physics, including dark matter decay and annihilation [15] , accreting or evaporating primordial black holes [16] , and primordial magnetic fields [17] , all predict increased gas temperatures and are unlikely to account for the observed amplitude. It is possible that some of these sources could also increase $T_R$ through mechanisms such as synchrotron emission associated with primordial black holes [18] or the relativistic electrons resulting from the decay of metastable particles [19] , but it is unclear if this could compensate for the increased gas temperature. Measurements from ARCADE-2 [20] suggest an isotropic radio background not explained by known source populations, but that interpretation has not reached consensus [21] and the sources would need to be present at z~20 to affect the radiation environment relevant to the observed signal.

While we have performed many tests to be confident that the observed profile is from a global absorption of the microwave background by hydrogen gas in the early Universe, we seek confirmation observations from other instruments. Several experiments similar to EDGES are underway. Those closest to achieving the performance required to verify the profile include LEDA [22], SCI-HI/PRIZM [23], and SARAS-2 [24]. More-sophisticated foreground models than those used in this analysis may lower the hardware performance requirements, as well as lead to better profile recovery since low-order modes in our fitted profile shape are degenerate with our foreground model and potentially under-constrained. Singular value decomposition of training sets constructed from simulated instrument error terms and foreground contributions can produce optimized basis sets for model fitting [25, 26]. We plan to apply these promising techniques to our data processing. The best measurement of the observed profile may ultimately be conducted in space, where the Earth's atmosphere and ionosphere will not influence the propagation of the astronomical signal, potentially reducing the burden of the foreground model. The measurement could be made from the lunar farside [27], in orbit or on the surface, exploiting the Moon as a shield to block FM radio signals and other Earth-based transmitters.

This result should bolster ongoing efforts to detect the statistical properties of spatial fluctuations in the 21 cm signal using interferometric arrays. It provides the first direct evidence that a signal exists for these telescopes to detect. The Hydrogen Epoch of Reionization Array [28] (HERA) is becoming operational over the next two years to characterize the power spectrum of redshifted 21 cm fluctuations between 100-200 MHz during the reionisation epoch, when the 21 cm signal is expected in emission. HERA plans to extend its operational band to 50 MHz over this time. It will likely have sufficient thermal sensitivity to detect any power spectrum signal associated with the observed profile, hence it may be first to validate the observed absorption signal. But hurdles remain as foregrounds have proven to be more challenging for interferometers than expected and sufficient foreground mitigation to detect the 21 cm power spectrum has yet to be demonstrated by any of the currently operating arrays, including LOFAR [29] and MWA [30]. Expansion of an existing Long Wavelength Array [31] station would provide the sensitivity to pursue power spectrum detection. When constructed, the planned SKA Low-Frequency Aperture Array (skatelescope.org) should be able to detect the power spectrum associated with the absorption profile and eventually image the 21 cm signal.

## REFERENCES

1. Pritchard, J. R. & Loeb, A., 21 cm cosmology in the 21st century. *Reports on Progress in Physics* **75** (8), 086901 (2012).

2. Cohen, A., Fialkov, A., Barkana, R. & Lotem, M., Charting the Parameter Space of the Global 21-cm Signal. *Mon. Not. R. Astron. Soc.* **472** (2), 1915-1931 (2017).

3. Barkana, R. Possible interaction between baryons and dark-matter particles revealed by the first stars, *Nature,* **555,** 71 (2018).

4. Monsalve, R. A., Rogers, A. E. E., Bowman, J. D. & Mozdzen, T. J., Results from EDGES High-Band: I. Constraints on Phenomenological Models for the Global 21 cm Signal. *Astrophys. J.* **847** (1), 64 (2017).

5. Alves, M. I. R. *et al.*, The HIPASS survey of the Galactic plane in radio recombination lines. *Mon. Not. R. Astron. Soc.* **450** (2), 2025-2042 (2015).

6. Rogers, A. E. E., Bowman, J. D., Vierinen, J., Monsalve, R. A. & Mozdzen, T., Radiometric Measurements of Electron Temperature and Opacity of Ionospheric Perturbations. *Rad. Sci.* **50** (2), 130-137 (2015).

7. Sokolowski, M. *et al.*, The Impact of the Ionosphere on Ground-based Detection of the Global Epoch of Reionization Signal. *Astrophys. J.* **813** (1), 18 (2015).

8. Zaldarriaga, M., Furlanetto, S. R. & Hernquist, L., 21 Centimeter Fluctuations from Cosmic Gas at High Redshifts. *Astrophys. J.* **608** (2), 622-635 (2004).

9. Ali-Haïmoud, Y. & Hirata, C. M., HyRec: A fast and highly accurate primordial hydrogen and helium recombination code. *Phys. Rev. D* **83** (4), 043513 (2011).

10. Shaw, J. R. & Chluba, J., Precise cosmological parameter estimation using COSMOREC. *Mon. Not. R. Astron. Soc.* **415** (2), 1343-1354 (2011).

11. Furlanetto, S. R., The global 21-centimeter background from high redshifts. *Mon. Not. R. Astron. Soc.* **371**, 867-878 (2006).

12. Planck Collaboration et al., Planck 2015 results XIII. Cosmological parameters. *Astron. Astrophys.* **594**, A13 (2016).

13. Chen, X. & Miralda-Escude, J., The Spin-Kinetic Temperature Coupling and the Heating Rate due to Lyα Scattering before Reionization: Predictions for 21 Centimeter Emission and Absorption. *Astrophys. J.* **602**, 1-11 (2004).

14. Tashiro, H., Kadota, K. & Silk, J., Effects of dark matter-baryon scattering on redshifted 21 cm signals. *Phys. Rev. D* **90**, 083522 (2014).

15. Lopez-Honorez, L., Mena, O., Moliné, Á., Palomares-Ruiz, S. & Vincent, A. C., The 21 cm signal and the interplay between dark matter annihilations and astrophysical processes. *JCAP* **2016** (8), 004 (2016).

16. Tashiro, H. & Sugiyama, N., The effect of primordial black holes on 21-cm fluctuations. *Mon. Not. R. Astron. Soc.* **435** (4), 3001-3008 (2013).

17. Schleicher, D. R. G., Banerjee, R. & Klessen, R. S., Influence of Primordial Magnetic Fields on 21 cm Emission. *Astrophys. J.* **692** (1), 236-245 (2009).

18. Biermann, P. L. *et al.*, Cosmic backgrounds due to the formation of the first generation of supermassive black holes. *Mon. Not. R. Astron. Soc.* **441**, 1147-1156 (2014).

19. Cline, J. M. & Vincent, A. C., Cosmological origin of anomalous radio background. *JCAP* **2013** (2) 011 (2013).

20. Seiffert, M. *et al.*, Interpretation of The Arcade 2 Absolute Sky Brightness Measurement. *Astrophys. J.* **734** (6) (2011).

21. Vernstrom, T., Norris, R. P., Scott, D. & Wall, J. V., The deep diffuse extragalactic radio sky at 1.75 GHz. *Mon. Not. R. Astron. Soc.* **447** (3), 2243-2260 (2015).

22. Bernardi, G. *et al.*, Bayesian constraints on the global 21-cm signal from the Cosmic Dawn. *Mon. Not. R. Astron. Soc.* **461** (3), 2847-2855 (2016).

23. Voytek, T. C., Natarajan, A., Jáuregui García, J. M., Peterson, J. B. & López-Cruz, O., Probing the Dark Ages at z ∼ 20: The SCI-HI 21 cm All-sky Spectrum Experiment. *Astrophys. J. Let.* **782** (1), L9 (2014).

24. Singh, S. *et al.*, First Results on the Epoch of Reionization from First Light with SARAS 2. *Astrophys. J. Let.* **845** (2), L12 (2017).

25. Switzer, E. R. & Liu, A., Erasing The Variable: Empirical Foreground Discovery for Global 21 cm Spectrum Experiments. *Astrophys. J.* **793** (102) (2014).

26. Vedantham, H. K. *et al.*, Chromatic effects in the 21 cm global signal from the cosmic dawn. *Mon. Not. R. Astron. Soc.* **437**, 1056-1069 (2014).

27. Burns, J. O. *et al.*, A Space-based Observational Strategy for Characterizing the First Stars and Galaxies Using the Redshifted 21 cm Global Spectrum. *Astrophys. J.* **844** (1), 33 (2017).

28. DeBoer, D. R. *et al.*, Hydrogen Epoch of Reionization Array (HERA). *PASP* **129** (974), 045001 (2017).

29. Patil, A. H. *et al.*, Upper Limits on the 21 cm Epoch of Reionization Power Spectrum from One Night with LOFAR. *Astrophys. J.* **838** (1), 65 (2017).

30. Beardsley, A. P. *et al.*, First Season MWA EoR Power spectrum Results at Redshift 7. *Astrophys. J.* **833** (1), 102 (2016).

## ACKNOWLEDGEMENTS

We thank CSIRO for providing site infrastructure and access to facilities. We thank the MRO Support Facility team, especially Michael Reay, Lou Puls, John Morris, Suzy Jackson, Brett Hiscock, and Kevin Ferguson. We thank Catherine Bowman, Delani Cele, Christopher Eckert, Leroy Johnson, Morgan Goodrich, Hamdi Mani, James Traffie, and Ken Wilson for instrument contributions. We thank Rennan Barkana for theory contributions and Gilbert Holder, Tanmay Vachaspati, Chris Hirata, and Jens Chluba for useful exchanges. We thank Colin Lonsdale, Heidi Johnson, Jacqueline Hewitt, and Jack Burns. We thank Carolyn and Mark Halleen for site and logistical support. This work was supported by the NSF through awards AST-0905990, AST-1207761, and AST-1609450. RAM acknowledges support from NASA Ames Research Center (NNX16AF59G) and the NASA Solar System Exploration Research Virtual Institute (80ARC017M0006). This scientific work makes use of the Murchison Radio-astronomy Observatory. We acknowledge the Wajarri Yamatji people as the traditional owners of the Observatory site.

## AUTHOR CONTRIBUTIONS

J. Bowman, R. Monsalve, and A. Rogers contributed to all activities. N. Mahesh and T. Mozdzen modelled instrument properties, performed laboratory calibrations, and contributed to preparation of this manuscript.

The authors have no competing financial interests.

Correspondence and requests for materials should be addressed to judd.bowman@asu.edu

**Table 1.  Sensitivity to possible calibration errors**

| Error Source | Estimated Uncertainty | Modeled Error Level | Recovered Amplitude (K) |
|---|---|---|---|
| LNA S11 magnitude | 0.1 dB | 1.0 dB | 0.51 |
| LNA S11 phase (delay) | 20 ps | 100 ps | 0.48 |
| Antenna S11 magnitude | 0.02 dB | 0.2 dB | 0.50 |
| Antenna S11 phase (delay) | 20 ps | 100 ps | 0.48 |
| No loss correction | N/A | N/A | 0.51 |
| No beam correction | N/A | N/A | 0.48 |

**TABLE AND FIGURE LEGENDS**

**Table 1. Sensitivity to possible calibration errors.** The estimated uncertainty for each case is based on empirical values from laboratory measurements and repeatability tests. Modeled error levels were chosen conservatively to be 5 or 10 times larger than the estimated uncertainty error levels.

**Figure 1. Summary of detection.** Panel (a) shows the measured spectrum for the reference dataset after filtering for data quality and RFI. The spectrum is dominated by Galactic synchrotron emission. The next two panels show residuals after fitting and removing the foreground model only (b) and the combined foreground and 21 cm models (c), respectively. Panel (d) shows the recovered 21 cm absorption model profile with SNR 37, amplitude 0.53 K, centre frequency 78.1 MHz, and width 18.7 MHz. Panel (e) presents the 21 cm model summed with its residuals from panel (c).

**Figure 2. Best-fit 21 cm absorption profiles for each hardware case.** Each profile is summed with its residuals and plotted against redshift and age of the Universe. The thick black line is the highest SNR=52 model fit that resulted from any of our six hardware configurations (see Methods) processed using 60-99 MHz and a 4-term polynomial foreground model. The thin lines are the best fits from each of the other hardware configuration cases, except for the profile that extends to z>26, which is reproduced from Figure 1 and used the same data as the SNR=52 fit but a different foreground model and the full band.

**METHODS**

**1. Instrument**

The EDGES experiment is located at the Murchison Radio-astronomy Observatory (MRO) in Western Australia (26.72 S, 116.61 E), which is the same radio-quiet [32] site used by the Australian SKA Precursor, the Murchison Widefield Array, and the planned SKA Low-Frequency Aperture Array. An early version of EDGES placed the first empirical lower limit on the duration of reionisation [33] . EDGES presently consists of three instruments: a high-band instrument [34, 4, 35] sensitive to 90-200 MHz (14>z>6) and two low-band instruments (low-1 and low-2) operating over 50-100 MHz (27>z>13). Each instrument yields spectra with 6.1 kHz resolution. In each instrument, sky radiation is collected by a wideband dipole-like antenna consisting of two rectangular metal panels mounted horizontally above a metal ground plane. Similar compact dipole antennas are used elsewhere in radio astronomy [36, 37] . A receiver is installed underneath the ground plane and a balun [38] is used to guide radiation from the antenna panels to the receiver. A rectangular shroud surrounds the base of the balun to shield the vertical currents in the balun tubes, which are strongest at the base. This lowers the gain toward the horizon due to the vertical currents.

Extended Data Figure 1 shows a block diagram of the system. A mechanical input switch at the front of the receiver allows the antenna to be connected to a remote Vector Network Analyzer (VNA) for accurate measurement of the antenna S11 or to be connected to the primary receiver path to measure the sky noise spectrum. When measuring the sky noise spectrum a second mechanical switch connects the LNA to the antenna or a 26 dB attenuator which acts as a load or a well matched noise source depending on the state of the electrical switch on the noise source. This performs the three-position switching operation [39] needed to provide the first stage of processing discussed below. After the LNA and post-amplifier, another noise source is used to inject noise below 45 MHz. This "out of band" conditioning improves the linearity and dynamic range of the analog-to-digital converter (ADC) needed for accurate cancellation of the receiver bandpass afforded by the three-position switching. A thermoelectric system maintains a constant temperature in the receiver in the field and in the laboratory. The system mitigates against RFI using designs and analysis strategies adapted from the Deuterium Array [40]. Similar approaches to instrument design are employed by SARAS-2 [41] and LEDA [42].

The low-band instrument design differs from the published descriptions of the high-band only by: 1) The addition of a 3 dB attenuator within the LNA before the PHEMT transistor at the input. The attenuator improves the LNA impedance match thereby reducing the sensitivity to measurement errors of the LNA and antenna reflection coefficients, especially errors in reflection phase, while adding only a small fraction of noise compared with the sky noise. Larger values of attenuation would begin to add significant noise at 100 MHz. 2) The use of a scaled antenna that is precisely double the size of the high-band antenna. 3) The use of a larger ground plane. Each low-band ground plane consists of a 2x2 meter solid metal central assembly surrounded by metal mesh that spans 30x30 meters, with the outer five meters shaped as saw-tooth perforated edges. Low-1 was initially operated with a 10x10 meter ground plane and later extended to full size. The full size 30x30 meter ground plane reduces the beam chromaticity and makes the beam less sensitive to conditions of the soil. Extended Data Figure 2 shows the low-1 and low-2 antennas, Extended Data Figure 3 plots the measured reflection coefficients, and Extended Data Figure 4 shows cuts through the antenna beam pattern model.

## 2. Calibration

We implement end-to-end absolute calibration for the low-band instruments following the techniques developed for the high-band [43, 34]. The calibration procedure involves taking reference spectra in the laboratory with the receiver connected to hot and ambient loads, as well as to open and shorted cables. Similar techniques are employed in other microwave measurements [44, 45]. S11 measurements using a VNA are acquired for the calibration sources and the LNA. The input connection to the receiver box provides the "reference plane" for all VNA measurements. In order to correct for the losses in the hot load used in the laboratory for calibration, full S-parameters are measured of the short cable from the heated resistor in the hot load. The accuracy of the reflection coefficient measurements is improved by accounting for the actual resistance of the VNA calibration 50 Ohm load and the added inductance in it due to skin effects [46] in the few mm of transmission line between the reference plane and internal load [47].

The calibration spectra and reflection coefficient measurements acquired in the laboratory are used to solve for free parameters [34] in equations that account for the impedance mismatches between the receiver and the antenna as well as the correlated and uncorrelated LNA noise waves. Laboratory calibration is performed with the receiver temperature controlled to the default 25°C, as well as at 15°C and 35°C in order to assess thermal dependence of the calibration parameters. Extended Data Figure 5 shows calibration parameter solutions for both receivers.

Following calibration in the laboratory, a test is made by measuring the spectrum of a ~300 K passive load with deliberate impedance mismatch that approximately mimics the reflection from the antenna in magnitude and phase. We call this device an "artificial antenna simulator". The reflection coefficient of the antenna simulator is measured and applied to yield calibrated integrated spectra. Extended Data Figure 3 shows measured antenna simulator reflection coefficients. Once corrected, the integrated spectra are expected to be spectrally flat, with a noise temperature that matches the physical temperature of the passive load. The flatness of the integrated spectra is quantified through the RMS of the residuals after removing a constant term. Typical residuals RMS are ~0.025 K across 50-100 MHz. If three polynomial terms are removed, the residuals decrease to ~0.015 K and are limited by integration time.

A second test of the calibration is made by measuring the spectrum of a noise source followed by a filter and 10 foot cable that adds about 30 ns of two-way delay. The device yields a spectrum similar in shape to the sky foreground with a strength of about 10,000 K (seven times larger than the typical sky temperature observed by EDGES) at 75 MHz and has a reflection coefficient of -6 dB in magnitude with phase slope similar to that of the antenna. Typical residuals are below 300 mK with five polynomial terms removed and are limited by integration time. Assuming any residuals scale with input power, this corresponds to 45 mK residuals at the typical observed sky temperature. This test is more sensitive than the passive simulator, especially to errors in the measurements of reflection coefficient as the signal is 33 times stronger than that of the ~300 K load and the magnitude of this simulator's reflection is higher than that of the passive simulator and of the real antenna.

Losses in the balun and losses due to the finite ground plane are corrected for during data processing using models. The balun loss model is validated against S-parameter measurements. Frequency-dependent beam effects are compensated for by modeling and subtracting spectral structure using electromagnetic (EM) beam models and a diffuse sky map template [35] . The nominal beam model accounts for the finite metal ground plane over soil with relative permittivity 3.5 and conductivity $2 \times 10^{-2}$ S/m [48] . The sky template is produced by extrapolating the 408 MHz all-sky radio map [49] to the observed frequencies using a spectral index in brightness temperature of -2.5 [43, 35] .

## 3.  Data and Processing

Examples of raw and processed data are shown in Extended Data Figure 6. Data processing occurs in three primary stages. In the first stage, three raw spectra that have been accumulated for 13 seconds each from the antenna input and two internal reference noise sources are converted to a single partially-calibrated spectrum [39] . Individual 6.1 kHz channels above a fixed power threshold are assigned zero weight to excise RFI. The threshold is normally set at three times the RMS of the residuals after the removal of a constant and a slope in a sliding 256 spectral channel window. Similarly, any partially-calibrated spectrum with average power above that expected from the sky or with large residuals is discarded. A weighted average of many successive spectra is taken, typically over several hours. Outlier channels after a Fourier series fit to the entire accumulated spectrum are again assigned zero weight. This second pass assigns zero weight to lower levels of RFI and broader RFI signals than the initial pass.

In the second stage of processing, the partially-calibrated spectra are fully calibrated using the calibration parameters from the laboratory and the antenna S11 measurements taken periodically in the field. Beam chromaticity corrections are applied after averaging the model over the same range of LST as in the spectra. The spectra are then corrected for the balun and ground plane loss and output with a typical smoothing to spectral bins with 390.6 kHz resolution.

In the third stage of processing, spectra for each LST block of several hours within each day are fit with a foreground model (see description of models below). An RMS value of the residuals is computed for each block and blocks above a selected threshold are discarded typically because of broadband RFI—or solar activity in daytime data—which were not detected in the earlier processing stages. A weighted average is then taken of the accepted blocks and a weighted least squares solution is made using a foreground model along with the model representing the 21 cm absorption signal. Extended Data Figure 7 plots the final weights for each spectral bin, equivalent to the RFI occupancy.

The observations used for the primary analysis presented in this work are from low-1 spanning 2016 day 252 through 2017 day 94 (configuration H2 below). The data are filtered to retain only local Galactic hour angles (GHA) from 6 to 18 hours. GHA is equivalent to LST offset by 17.75 hours.

## 4. Parameter Estimation

The foreground polynomial used for the analysis presented in Figure 1 is physically motivated, with five terms based on the known spectral properties of the Galactic synchrotron spectrum and Earth's ionosphere [6, 50] . It is given by:

$$T_F(\upsilon) \approx a_0 \ (\upsilon/\upsilon_c)^{-2.5} + a_1 \ (\upsilon/\upsilon_c)^{-2.5} \log(\upsilon/\upsilon_c) + a_2 \ (\upsilon/\upsilon_c)^{-2.5} [\log(\upsilon/\upsilon_c)]^2 + a_3 \ (\upsilon/\upsilon_c)^{-4.5} + a_4 \ (\upsilon/\upsilon_c)^{-2} . \quad (2)$$

Here $T_F(\upsilon)$ is the brightness temperature of the foreground emission, $\upsilon$ is frequency, and the $a_n$ coefficients are fit to the data. The above function is a linear approximation to:

$$T_F(\upsilon) = b_0 \ (\upsilon/\upsilon_c)^{-2.5+b_1+b_2 \log(\upsilon/\upsilon_c)} \ e^{-b_3(\upsilon/\upsilon_c)^{-2}} + b_4(\upsilon/\upsilon_c)^{-2}, \quad (3)$$

which is directly connected to the physics of the foreground and ionosphere. The factor of -2.5 in the first exponent is the typical foreground power-law spectral index, $b_0$ is an overall foreground scale factor, $b_1$ allows for a correction to the typical foreground spectral index (which varies by ~0.1 across the sky), and $b_2$ captures any contributions from a higher-order foreground spectral term [51, 52]. Ionospheric contributions are contained in $b_3$ and $b_4$, which allow for the ionospheric absorption of the foreground and emission from hot electrons in the ionosphere, respectively. This model is also able to partially capture some instrumental effects, such as additional spectral structure from chromatic beams or small errors in calibration.

We also use a more-general polynomial model in many of our trials that enables us to explore signal recovery with varying numbers of polynomial terms. This model is given by:

$$T_F(\upsilon) = \sum_{n=0}^{N-1} a_n \, \upsilon^{n-2.5},$$ (4)

where $N$ is the number of terms and the $a_n$ coefficients are again fit to the data. As with the physical model, the -2.5 index in the exponent makes it easier for the model to match the foreground spectrum. Both foreground models yield consistent absorption profile results.

The 21 cm absorption profile is modelled as a flattened Gaussian shape, given by:

$$T_{21}(\upsilon) = -A \, \left( \frac{1 - e^{-\tau \, e^B}}{1 - e^{-\tau}} \right),$$ (5)

where

$$B = \frac{4 \, (\upsilon - \upsilon_0)^2}{w^2} \quad \ln\left[ -\ln\left( \frac{1 + e^{-\tau}}{2} \right) \middle/ \tau \right]$$ (6)

and $A$ is the absorption amplitude, $\upsilon_0$ is the centre frequency, $w$ is the full width at half maximum, and $\tau$ is a flattening factor. This model is not a description of the physics that creates the 21 cm absorption profile, but rather is a suitable functional form to capture the basic shape of the profile. Extended Data Figure 8 shows the best-fit profile model and residuals to the model from fits by the two foreground models.

We report parameter fits from a gridded search over the $\upsilon_0$, $w$, and $\tau$ parameters in the 21 cm model. For each step in the grid, we conducted a linear weighted least squares fit, solving simultaneously for the foreground coefficients and the absorption profile amplitude. The best-fit absorption profile model maximizes the SNR in the gridded search. The amplitude fit uncertainty accounts for covariance between the foreground coefficients and the profile amplitude, as well as for noise.

Fitting both foreground and 21 cm models simultaneously yields residuals that decrease with integration time with an approximately noise-like $(1/\sqrt{t})$ trend for the duration of the observation, whereas fitting only the foreground model yields residuals that decrease with time initially for the first ~10% of the integration and then saturate, as shown in Extended Data Figure 9.

We also performed a Monte Carlo Markov Chain analysis, shown in Extended Data Figure 10, for the H2 case using a 5-term polynomial foreground model and a subset of the band covering 60-94 MHz. The amplitude parameter is most covariant with the flattening. The 99% statistical confidence intervals on the four 21 cm model parameters are: $A = 0.52^{+0.42}_{-0.18}$ K, $\upsilon_0 = 78.3^{+0.2}_{-0.3}$ MHz, $w = 20.7^{+0.8}_{-0.7}$ MHz, and $\tau = 6.5^{+5.6}_{-2.5}$. These intervals do not include any systematic error from differences across the hardware configurations and processing trials. When the flattening parameter is fixed to $\tau = 7$, statistical uncertainty in the 21 cm model amplitude fit is reduced to approximately ±0.02 K. Extended Data Table 1 shows that the various hardware configurations and processing trials with fixed $\tau = 7$ yield best-fit parameter ranges spanning: $0.37 < A < 0.67$ K, $77.4 < \upsilon_0 < 78.5$ MHz, and $17.0 < w < 22.8$ MHz. This systematic variation is likely due to the limited data in the some of the configurations, small calibration errors, residual chromatic beam effects, and potentially to structure in the Galactic foreground that increases when the Galactic plane is overhead. For each parameter, taking the outer bounds of the statistical confidence ranges from the H2 comprehensive MCMC analysis and the best-fit variations between validation trials in Extended Table 1 yields our estimate of the 99% confidence intervals that we reported in the main article.

## 5. Verification Tests

Here we list the tests we performed to verify the detection. The absorption profile is detected from data obtained in the following hardware configurations:

H1.  Low-1 with 10x10 meter ground plane
H2.  Low-1 with 30x30 meter ground plane
H3.  Low-1 with 30x30 meter ground plane and recalibrated receiver
H4.  Low-2 with north-south dipole orientation
H5.  Low-2 with east-west dipole orientation
H6.  Low-2 with east-west dipole orientation and balun shield removed to check for any resonance that might result from a slot antenna being formed in the joint between the two halves of the shield.

The absorption profile is detected in data processed with the following configurations:

P1.  All hardware cases using two independent processing pipelines
P2.  All hardware cases divided into temporal subsets
P3.  All hardware cases with chromatic beam corrections on/off
P4.  All hardware cases with ground loss and balun loss corrections on/off
P5.  All hardware cases calibrated with four different antenna S11 measurements

P6.   All hardware cases using 4-term foreground model (Eqn. 4) over frequency range 60-99 MHz

P7.   All hardware cases using 5-term foreground model (Eqn. 4) over frequency range 60-99 MHz

P8.   All hardware cases using physical foreground model (Eqn. 2) over frequency range 51-99 MHz

P9.   All hardware configurations using various additional combinations of 4-, 5-, and/or 6-term foreground models and frequency ranges

P10.  Case H2 binned by LST/GHA

P11.  Case H2 binned by UTC

P12.  Case H2 binned by buried conduit temperature as a proxy for the ambient temperature at the receiver and the temperature of the cable that connects the receiver frontend under the antenna to the backend in the control hut

P13.  Case H2 binned by Sun above/below horizon

P14.  Case H2 binned by Moon above/below horizon

P15.  Case H2 with added Galaxy up/down differencing calibration

P16.  Case H2 calibrated with low-2 solutions

P17.  Case H4 calibrated with laboratory measurements at 15 and 35°C

P18.  Cases H2-H3 calibrated with laboratory measurements spanning two years

Extended Data Table 1 lists the profile properties from each of the hardware configurations with the standard processing (P6) and Figure 2 illustrates the corresponding best-fit profiles. The variations in best-fit SNR between the configurations are largely explained by differences in total integration times for each configuration, except for H1 which was limited by its ground plane performance. We acquired the most data in configuration H1, with approximately 11 months of observations, followed by H2 with six months. The other configurations were each operated for 1-2 months before the analysis presented here. Extended Data Table 2 lists the profile amplitudes for data binned by GHA for both processing pipelines.

The following additional verification tests were performed to check specific aspects of the instrument, laboratory calibration, and processing pipelines, including:

- We processed simulated data and recovered injected profiles.
- We searched for a similar profile at the scaled frequencies in high-band data and found no corresponding profile.
- We measured the antenna S11 of low-2 with the VNA connected to its receiver with a short two-meter cable and found nearly identical results as with the 100-meter cable used in operations.
- We acquired *in situ* S11 measurements that matched our model predictions of the low-2 balun with the antenna terminal shorted and open. This was done to verify our model for the balun loss.
- We tested the performance of the receivers in the laboratory using artificial antenna sources connected directly to the receivers, as described above.
- We cross-checked our beam models using three electromagnetic numerical solvers: CST, FEKO, and HFSS. Although no beam model is required to detect the profile because the EDGES antenna is designed to be largely achromatic, we performed the cross-check since we apply beam corrections in the primary analysis.

## 6. Sensitivity to Systematic Errors

In this section, we discuss in more detail several primary categories of potential systematic errors and the validation steps we performed.

### 6.1. Beam and Sky Effects

Beam chromaticity is larger than can be accounted for with EM models of the antenna on an infinite ground plane [53]. For both ground plane sizes, the RMS of residuals to low-order foreground polynomial model fits of data matched EM modeling when the model accounted for the finite ground plane size and included the effects of the dielectric constant and conductivity of the soil under the ground plane. The residual structures themselves matched qualitatively.

Comparing EM solvers for beam models, we found that for infinite ground plane models, the change in the absolute gain of the beam with frequency at every viewing angle (theta, phi) was within ±0.006 between solvers and that residuals after foreground fits to simulated spectra were within a factor of two. For models with finite ground planes and real soil properties, we found that correcting H1 data using beam models from FEKO and HFSS in integral solver modes resulted in nearly identical 21 cm model parameter values, although using an HFSS model for the larger ground plane in H2 resulted in a lower-SNR fit to the profile than a FEKO model, but still higher-SNR than no beam correction (see Extended Data Table 1).

The low-2 instrument was deployed 100 meters west of the control hut, compared to 50 meters east of the hut for low-1. In the east-west antenna orientation, the low-2 dipole response null was aimed approximately at the control hut and the beam pattern on the sky was rotated compared to north-south. Obtaining consistent absorption profiles with the two sizes of low-1 ground planes (H1 vs. H2/H3) and with both low-2 antenna orientations (H4 vs. H5/H6) suggests that beam effects are not responsible for the profile, while obtaining the same results from both low-2 antenna orientations also disfavors polarized sky emission as a possible source of the profile. Obtaining consistent absorption profiles with the low-1 and low-2 instruments at different distances from the control hut and with both low-2 antenna orientations suggests that it is unlikely that the observed profile is produced by reflections of sky noise from the control hut or other surrounding objects or caused by RFI from the hut. Our understanding of hut reflections is further validated by the appearance of small sinusoidal ripples following 9-term foreground removal from low-1 spectra at GHA 20. These ripples are consistent with models of hut reflections and not evident at other GHA or in low-2 data.

### 6.2. Gain and Loss Errors

Many possible instrumental systematic errors and atmospheric effects that could potentially mimic the observed absorption profile are due to inaccurate or unaccounted for gains or losses in the propagation path within the instrument or Earth's atmosphere. If present, these effects would be proportional to the total sky noise power entering the system. The total sky noise power received by EDGES varies by a factor of three over GHA. If the observed absorption profile were due to gain or loss errors, the amplitude of the profile would be expected to vary in GHA proportional to the sky noise.

We tested for these errors by fitting for the absorption profile in observations binned by GHA in 4-hour and 6-hour blocks using H2 data. The test is complicated by the increase in chromatic beam effects in spectra when the sky noise power is large due to the presence of the Galactic plane in the antenna beam. We compensated for this by increasing the foreground model to up to six polynomial terms for the GHA analysis and using the FEKO antenna beam model to correct for beam effects. As evident in Extended Data Table 2, the best-fit amplitudes averaged over each GHA bin are consistent within the reported uncertainties and exhibit no substantial correlation with sky noise power. The same test performed using only a 4-term polynomial foreground model did yield variations with GHA, as did tests performed on data from low-1 with the 10x10 meter ground plane. We attribute the failure of these two cases to beam effects and possible foreground structure. Other cases tested had insufficient data for conclusive results, but did not show correlation with the total sky noise power.

The artificial antenna measurements described in the calibration section above provide verification of the smooth passband of the receiver after calibration. Since we observe the 0.5 K signature for all foreground conditions, including low foregrounds of ~1500 K at ~78 MHz, if the observed profile were an instrumental artifact due to an error in gain of the receiver, we would expect to see a scaled version of the profile with an amplitude of 0.5 K x (300 K / 1,500 K) = 0.1 K, when measuring the ~300 K artificial antenna. Instead, we see a smooth spectrum structure at the ~ 0.025 K level. With the 10,000 K artificial antenna, we would expect to see a 3.3 K profile yielding 0.5 K residuals after removing a five-term polynomial fit. Instead we find residuals that are less than 0.3 K.

Receiver calibration errors are disfavored as the source of the observed profile. Three verification tests were made to specifically investigate this possibility by processing data with inaccurate calibration parameter solutions. In verification test P18 we processed H2 and H3 datasets using each of the three low-1 receiver calibration solutions shown in the left column of Extended Data Figure 5. The observed profile was detected in each case, indicating the detection is robust to these small drifts in the calibration parameters over the two-year period spanning the use of the low-1 receiver. Second, in verification test P17 we processed H4 observations using the calibration solutions derived from laboratory measurements acquired with the receiver temperature held at both 15 and 35°C, even though it was controlled to 25°C for all observations. The profile was recovered even for these larger calibration differences, thus we infer the detection is robust to the much smaller ~0.1°C typical variations in the receiver temperature around its set point during operation. As a final check of the receiver properties, in P16 we calibrated the H2 dataset from low-1 using the receiver calibration solutions derived for low-2. The profile was recovered using a foreground polynomial model of seven terms over 53-99 MHz. This provides evidence that both receivers have generally similar properties and spectrally smooth responses, otherwise we would not expect the calibration solutions to be interchangeable in this manner.

## 6.3. RFI and FM Radio

RFI is found to be minimal in EDGES low-band measurements. We rule-out locally-generated broadband RFI from the control hut and a nearby ASKAP dish (>150 meters away) as the source of the profile because of the consistent profiles observed by both instruments and both low-2 antenna orientations, as noted above. There are no licensed digital TV transmitters in Australia below 174 MHz (see: www.acma.gov.au, ITU RCC-06). We have analyzed observations and rule-out the FM radio band, which spans 87.5 to 108 MHz, as the cause of the high-frequency edge of the observed profile. FM transmitters within ~3000 km of the MRO could be scattered from airplanes or meteors that burn up at an altitude of about 100 km in the mesosphere. Careful inspection of channels excised by our RFI detection algorithms and spectral residuals using the instrument's raw 6.1 kHz resolution, which oversamples the minimum 50 kHz spacing of the FM channel centres, shows that these signals are sparse and transient and show up after excision as mostly zero-weighted channels. More-persistent worldwide FM signals reflected from the Moon have been measured [54] from the MRO with flux density ~100 Jy. We find evidence for a sharp step of ~0.05 K at 87.5 MHz in our binned spectra when the Moon is above 45° elevation, which can be eliminated by using only data from when the Moon is below the horizon.

## 7. Atmospheric Molecular Lines

Atmospheric nitrous oxide line absorption was modeled using the JPL catalog line strength of $10^{-12.7}$ nm$^2$ MHz at 300 K and an abundance of 70 parts per billion. We assumed a 3000 K sky noise temperature and a line-of-sight path through the atmosphere at 8° elevation and integrated over the altitude span from 10 to 120 km. We find up to 0.001 mK absorption per line. With approximately 100 individual lines between 50-100 MHz, we conservatively estimate a maximum possible contribution of 0.1 mK.

## 8. Gas Temperature and Residual Ionization Fraction

For the gas thermal calculations, we used CosmoRec [10] to model the evolution of the electron temperature and residual ionization fraction for z<3000. We verified the output against solutions to equations [55] for the dominant contributions to the electron temperature evolution of adiabatic expansion and Compton scattering. We assume the gas temperature is in equilibrium with the electron temperature. To determine the residual ionization fraction required to produce sufficiently cold gas to account for the observed profile amplitude, we modeled a partial ionization step function in redshift. We used the ionization fraction from CosmoRec for redshifts above the transition and a constant final ionization fraction below the transition. We performed a grid search in transition redshift and final ionization fraction to identify the lowest transition redshift for the largest final ionization fraction that resulted in the required gas temperature. We found that a final ionization fraction of ~3 x $10^{-5}$ reached by z~500 would be sufficient to produce the required gas temperature. This is nearly an order of magnitude lower than the expected ~2 x $10^{-4}$ ionization fraction at similar ages from CosmoRec.

## Data Availability

The data that support the findings of this study are available from the corresponding author upon reasonable request.

## Code Availability

The code that supported the findings of this study are available from the corresponding author upon reasonable request.

## REFERENCES (METHODS)

31. Dowell, J., Taylor, G. B., Schinzel, F. K., Kassim, N. E. & Stovall, K., The LWA1 Low Frequency Sky Survey. *Mon. Not. R. Astron. Soc.* **469** (4), 4537-4550 (2017).

32. Bowman, J. & Rogers, A. E. E., VHF-band RFI in Geographically Remote Areas. *Proceedings of the RFI Mitigation Workshop. 29-31 March 2010. Groningen, the Netherlands* **POS**, Published online at http://pos.sissa.it/cgi-bin/reader/conf.cgi?confid=107, id.30 (2010).

33. Bowman, J. D. & Rogers, A. E. E., Lower Limit of dz>0.06 for the duration of the reionization epoch. *Nature* **468** (7325), 796-798 (2010).

34. Monsalve, R. A., Rogers, A. E. E., Bowman, J. D. & Mozdzen, T. J., Calibration of the EDGES High-band Receiver to Observe the Global 21 cm Signature from the Epoch of Reionization. *Astrophys. J.* **835** (1), 3 (2017).

35. Mozdzen, T. J., Bowman, J. D., Monsalve, R. A. & Rogers, A. E. E., Improved measurement of the spectral index of the diffuse radio background between 90 and 190 MHz. *Mon. Not. R. Astron. Soc.* **464** (4), 4995-5002 (2017).

36. Raghunathan, A., Shankar, N. U. & and Subrahmanyan, R., An Octave Bandwidth Frequency Independent Dipole Antenna. *IEEE Transactions on Anennas and Propagation* **61** (7), 3411-3419 (2013).

37. Ellingson, S. W., Antennas for the Next Generation of Low-Frequency Radio Telescopes. *IEEE Transactions on Antennas and Propagation* **53** (8), 2480-2489 (2005).

38. Roberts, W. K., A New Wide-Band Balun. *Proc. of the IRE* **45**, 1628-1631 (1957).

39. Bowman, J. D., Rogers, A. E. E. & Hewitt, J. N., Toward Empirical Constraints on the Global Redshifted 21 cm Brightness Temperature During the Epoch of Reionization. *Astrophys. J.* **676** (1), 1 (2008).

40. Rogers, A. E. E., Pratap, P., Carter, J. C. & Diaz, M. A., Radio frequency interference shielding and mitigation techniques for a sensitive search for the 327 MHz line of deuterium. *Radio Science* **40** (5), RS5S17 (2005).

41. Singh, S. *et al.*, SARAS 2: A Spectral Radiometer for probing Cosmic Dawn and the Epoch of Reionization through detection of the global 21 cm signal. Preprint at https://arxiv.org/abs/1710.01101 (2017).

42. Price, D. C. *et al.*, Design and characterization of the Large-Aperture Experiment to Detect the Dark Age (LEDA) radiometer systems. Preprint at https://arxiv.org/abs/1709.09313 (2017).

43. Rogers, A. E. E. & Bowman, J. D., Absolute calibration of a wideband antenna and spectrometer for accurate sky noise temperature measurement. *Rad. Sci.* **47** (RS0K06), 9 (2012).

44. Hu, R. & Weinreb, S., A Novel Wide-Band Noise-Parameter Measurement. *IEEE Transactions on Microwave Theory and Techniques* **52** (5), 1498-1507 (2004).

45. Belostotski, L., A Calibration Method for RF and Microwave Noise Sources. *IEEE Transactions on Microwave Theory and Techniques* **59** (1), 178-187 (2011).

46. Ramo, S. & Whinnery, J. R., *Fields and Waves in Modern Radio* (Wiley, New York, 1953).

47. Monsalve, R. A., Rogers, A. E. E., Mozdzen, T. J. & Bowman, J. D., One-Port Direct/Reverse Method for Characterizing VNA Calibration Standards. *ITMTT* **64** (8), 2631-2639 (2016).

48. Sutinjo, A. T. *et al.*, Characterization of a Low-Frequency Radio Astronomy Prototype Array in Western Australia. *IEEE Trans. Antennas Propag.* **63** (12), 5433-5442 (2015).

49. Haslam, C. G. T., Salter, C. J., Stoffel, H. & E., W. W., A 408 MHz all-sky continuum survey. II - The atlas of contour maps. *A&AS* **47**, 1, 2, 4-51, 53-142 (1982).

50. Chandrasekhar, S., *Radiative Transfer* (Courier Dover, New York, 1960).

51. de Oliveira-Costa, A. *et al.*, A model of diffuse Galactic radio emission from 10 MHz to 100 GHz. *Mon. Not. R. Astron. Soc.* **388** (1), 247-260 (2008).

52. Bernardi, G., McQuinn, M. & Greenhill, L. J., Foreground Model and Antenna Calibration Errors in the Measurement of the Sky-averaged λ21 cm Signal at z~ 20. *Astrophys. J.* **799** (1), 90 (2015).

53. Mozdzen, T. J., Bowman, J. D., Monsalve, R. A. & Rogers, A. E. E., Limits on foreground subtraction from chromatic beam effects in global redshifted 21 cm measurements. *Mon. Not. R. Astron. Soc.* **455** (4), 3890-3900 (2016).

54. McKinley, B. *et al.*, Low-frequency Observations of the Moon with the Murchison Widefield Array. *Astron. J.* **145** (1), 23 (2013).

55. Seager, S., Sasselov, D. D. & Scott, D., How Exactly Did the Universe Become Neutral? *Astrophys. J. Sup.* **128** (2), 407-430 (2000).

## EXTENDED DATA TITLES AND LEGENDS

**Extended Data Table 1. Best-fit parameter values for the 21 cm absorption profile for representative verification tests.** Model fits were performed by grid search with fixed τ=7. Sky time is the amount of time spent by the receiver in the antenna switch state and is 33% of wall-clock time. The data acquisition system has a duty cycle of ~50% and a spectral window function efficiency of ~50%, yielding effective integration times that are a factor of four smaller than the listed sky times.

**Extended Data Table 2. Recovered 21 cm profile amplitudes for various GHA.** Each block is centred on the GHA listed. The 6-hour bins used the 5-term physical foreground model fit simultaneously with the 21 cm profile amplitude between 64-94 MHz. The 4-hour bins used a 6-term polynomial foreground model fit between 65-95 MHz. All data are from hardware configuration H2. Sky temperatures are reported at 78 MHz.

**Extended Data Figure 1. Block diagram of low-band system.** The inset images show: (a) the capacitive tuning bar that feeds the dipoles at the top of the balun, (b) the SMA connector at the bottom of the balun coaxial transmission line where the receiver connects, (c) the low-1 receiver installed under its antenna with the ground plane cover plate removed, and (d) the inside of the low-1 receiver. The LNA is contained in the secondary metal enclosure in the lower-left corner of the receiver.

**Extended Data Figure 2. Low-band antennas.** Panel (a) shows the low-1 antenna with the 30x30 meter mesh ground plane. The darker inner square is the original 10x10 meter mesh. The control hut is 50 meters from the antenna. Panel (b) shows a close view of the low-2 antenna. The two elevated metal panels form the dipole-based antenna and are supported by fiberglass legs. The balun consists of the two vertical brass tubes in the middle of the antenna. The balun shield is the shoebox-sized metal shroud around the bottom of the balun. The receiver is under the white metal platform and not visible.

**Extended Data Figure 3. Antenna and simulator reflection coefficients.** In panels (a) and (b), measurements are plotted for hardware cases H2 (blue), H4 (red), and H6 (orange). The antennas are identically designed (except H6 has the balun shield removed), but are tuned manually during installation by adjusting the panel separation and the height of the small metal plate that connects one panel to the centre conductor of the balun transmission line on the other. The measurements were acquired *in situ*. In panels (c) and (d), the red curve is the 10,000 K artificial antenna noise source and the blue curve is the 300 K mismatched load.

**Extended Data Figure 4. Antenna beam model.** Panel (a) shows beam cross-sections in the E-plane (solid) and H-plane (dash) from FEKO for the H2 antenna and ground plane over soil. Cross-sections are plotted at: 50 (red), 70 (green), and 100 MHz (blue). Panel (b) shows the frequency-dependence of the gain at $\Theta = 0°$ (solid) and the 3 dB points at 70 MHz in the E-plane (dash) and H-plane (dot). Small undulations with frequency are shown in panel (c) after a 5-term polynomial has been removed from each of the curves. Simulated observations with this model yields 0.015 K (0.001%) residuals to a 5-term fit over frequency range 52-97 MHz at GHA=10 and 0.1 K (0.002%) residuals at GHA=0, showing the cumulative beam yields less chromaticity than the ~1% variations of the individual points plotted.

**Extended Data Figure 5. Calibration parameter solutions.** Panels (a) through (g) show the calibration parameter solutions for the low-1 receiver at its fixed 25°C operating temperature. It was calibrated on three occasions spanning two years, bracketing all of the low-1 observations reported. The first calibration was in August 2015 before commencing cases H1 and H2 (solid), the next was in May 2017 before H3 (dot), and the final was in September 2017 after the conclusion of H3 (dash). Panels (h) through (n) show the solutions for the low-2 receiver controlled to three different temperatures: 15°C (blue), 25°C (black), and 35°C (red).

**Extended Data Figure 6. Raw and processed spectra.** Panel (a) shows typical raw 13-second spectra from H2 for each of the receiver's internal "three-position" switch states. The small spikes on the right of the antenna spectrum are FM stations. In panel (b), the spectrum has been partially calibrated ($T_{3-pos}$) using the three raw spectra to correct gain and offset contributions in the receiver and cables, then fully calibrated ($T_{cal}$) by applying the calibration parameter solutions from the laboratory to yield the sky temperature. Panel (c) shows residuals to a fit of the fully calibrated spectrum with the physical foreground model.

**Extended Data Figure 7. Normalized channel weights.** The fraction of data integrated for each 390.6 kHz spectral bin. In panel (a), the FM band causes the low weights above 87 MHz because many 6.1 kHz raw spectral channels in this region are excised for all times. The weights are nearly identical across all hardware cases (H1-H6). Panel (b)

provides a close-up to show the weights below the FM band, where there is little RFI to excise.

**Extended Data Figure 8.  Residuals to 21 cm profile model.**  The black curve shows the best-fit 21cm profile model derived from the observations.  The blue and orange solid curves show fits to the model profile using the physical and 5-term polynomial foreground models, respectively.  The dashed lines show the residuals after subtracting the fits from the model.  These residuals are similar to those found when fitting the observations using only a foreground model, as shown in Figure 1 panel (b).

**Extended Data Figure 9.  Residual RMS as a function of integration time.**  The curves show the residual RMS after a best-fit model is removed at each integration time for the H2 dataset.

**Extended Data Figure 10.  Parameter estimation.**  Likelihood distributions for the foreground and 21 cm model parameters are shown for the H2 dataset.  Contours are drawn at the 68% and 95% probability levels. The foreground polynomial coefficients ($a_n$) are highly correlated with each other, while the 21 cm model parameters are largely uncorrelated except for the profile amplitude ($A$) and flattening ($\tau$).  Systematic uncertainties from the verification hardware cases are not represented here.

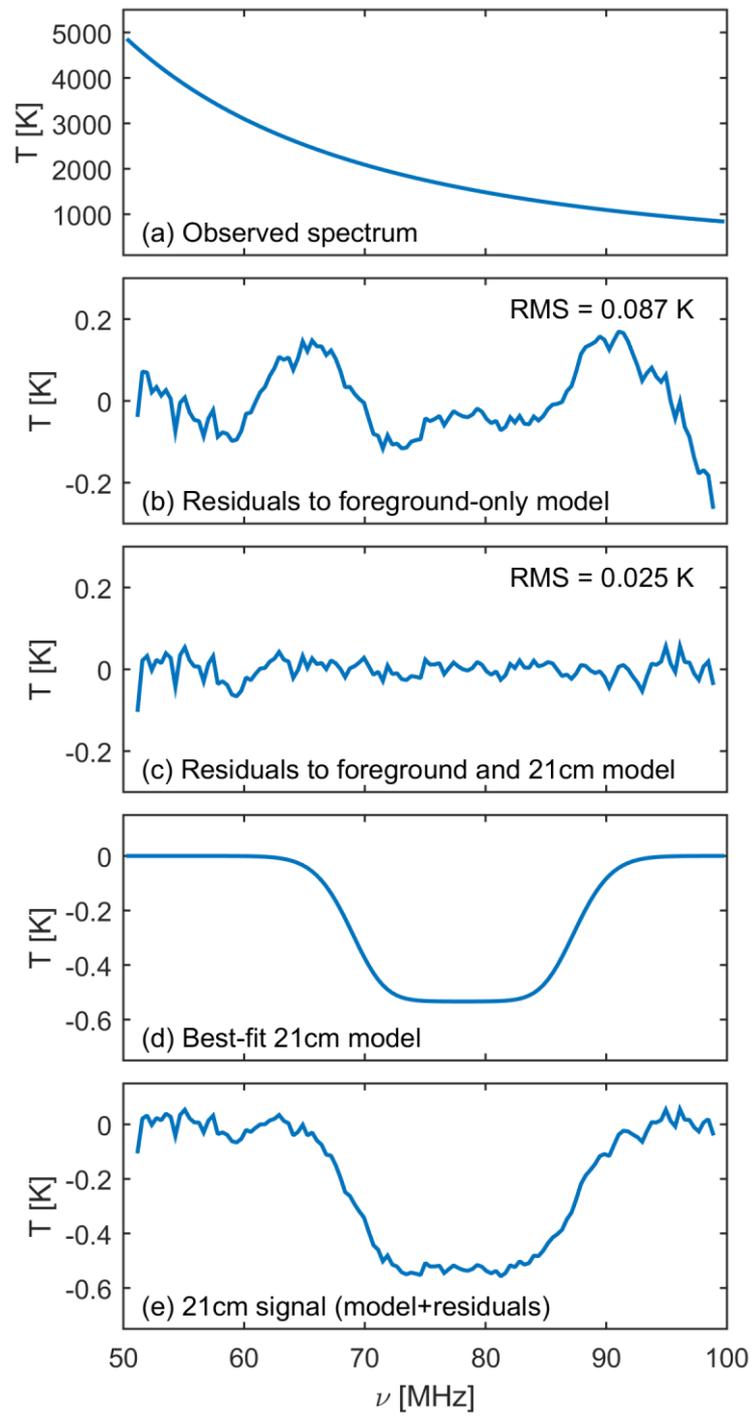

**Figure 1**

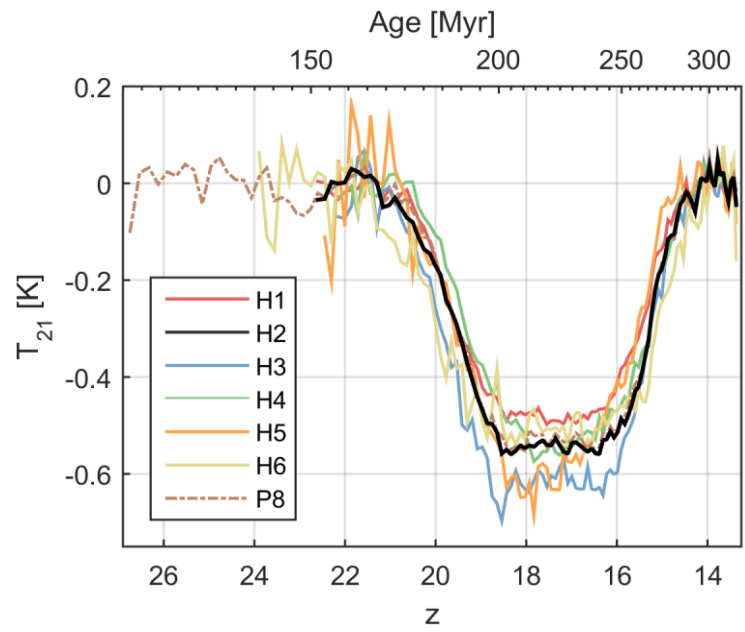

**Figure 2**

**Extended Data Table 1.  Best-fit parameter values for the 21 cm absorption profile for representative verification tests.**

| Configuration | Sky Time (hours) | SNR | Centre Frequency (MHz) | Width (MHz) | Amplitude (K) |
|---|---|---|---|---|---|
| **Hardware configurations (all P6)** | | | | | |
| H1 – low-1 10x10 ground plane | 528 | 30 | 78.1 | 20.4 | 0.48 |
| H2 – low-1 30x30 ground plane | 428 | 52 | 78.1 | 18.8 | 0.54 |
| H3 – low-1 recalibrated receiver | 64 | 13 | 77.4 | 19.3 | 0.43 |
| H4 – low-2 NS | 228 | 33 | 78.5 | 18.0 | 0.52 |
| H5 – low-2 EW | 68 | 19 | 77.4 | 17.0 | 0.57 |
| H6 – low-2 EW no balun shield | 27 | 15 | 78.1 | 21.9 | 0.50 |
| **Processing configurations (all H2 except P17)** | | | | | |
| P3  – No beam correction | | 19 | 78.5 | 20.8 | 0.37 |
| No beam correction (65-95 MHz) | | 25 | 78.5 | 18.6 | 0.47 |
| HFSS beam model | | 34 | 78.5 | 20.8 | 0.67 |
| FEKO beam model | | 48 | 78.1 | 18.8 | 0.50 |
| P4  – No loss corrections | | 25 | 77.4 | 18.6 | 0.44 |
| P7  – 5-term foreground polynomial (60-99 MHz) | | 21 | 78.1 | 19.2 | 0.47 |
| P8  – Physical foreground model (51-99 MHz) | | 37 | 78.1 | 18.7 | 0.53 |
| P14 – Moon above horizon | | 44 | 78.1 | 18.8 | 0.52 |
| Moon below horizon | | 40 | 78.5 | 18.7 | 0.47 |
| P17 – 15°C calibration (61-99 MHz, 5-term) | | 25 | 78.5 | 22.8 | 0.64 |
| 35°C calibration (61-99 MHz, 5-term) | | 16 | 78.9 | 22.7 | 0.48 |

**Extended Data Table 2. Recovered 21 cm profile amplitudes for various GHA**

| Galactic Hour Angle (GHA) | SNR | Amplitude (K) | Sky Temperature (K) |
|---|---|---|---|
| 6-hour bins | | | |
| 0 | 8 | 0.48 | 3999 |
| 6 | 11 | 0.57 | 2035 |
| 12 | 23 | 0.50 | 1521 |
| 18 | 15 | 0.60 | 2340 |
| 4-hour bins | | | |
| 0 | 5 | 0.45 | 4108 |
| 4 | 9 | 0.46 | 2775 |
| 8 | 13 | 0.44 | 1480 |
| 12 | 21 | 0.57 | 1497 |
| 16 | 11 | 0.59 | 1803 |
| 20 | 9 | 0.66 | 3052 |

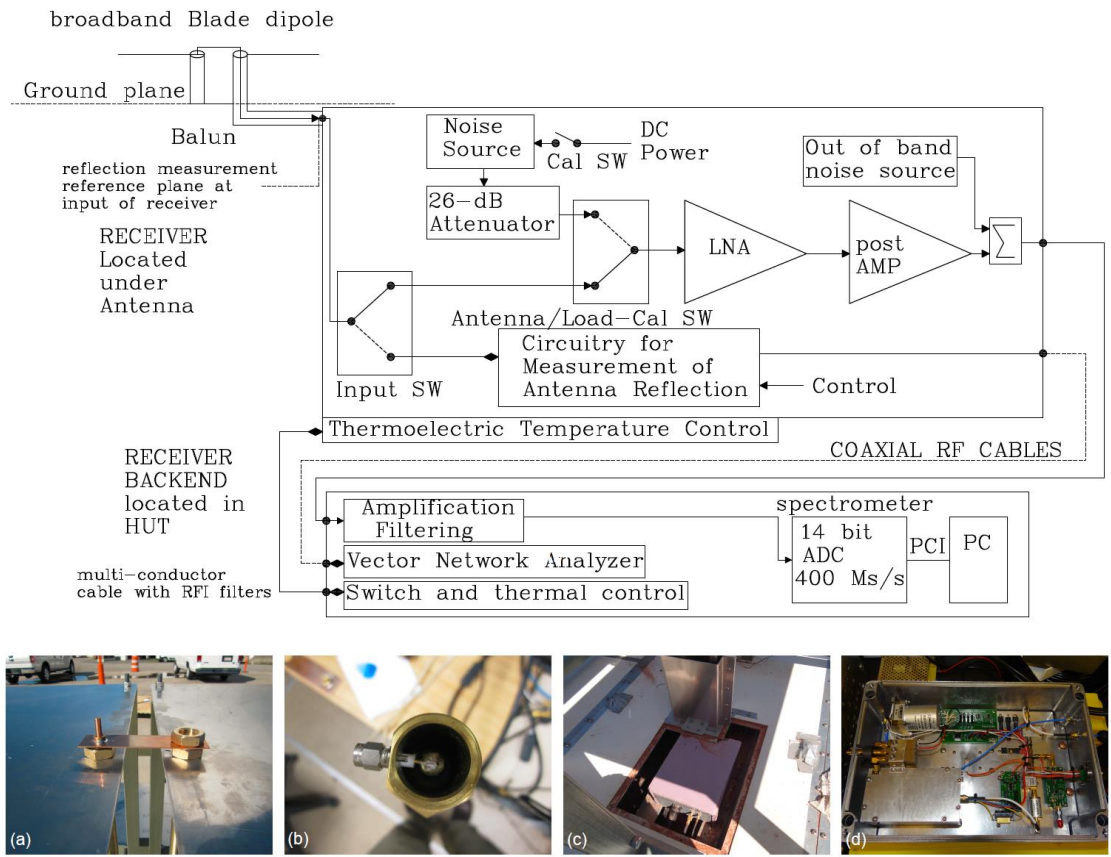

**Extended Data Figure 1**

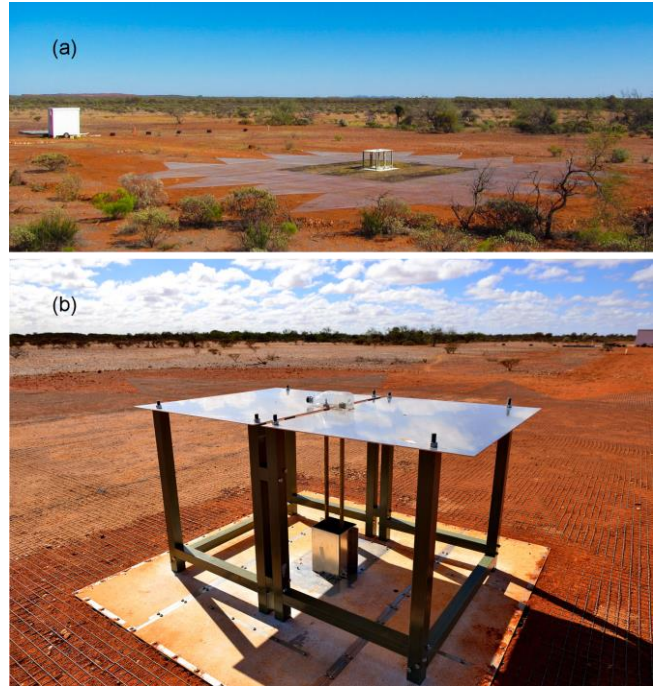

**Extended Data Figure 2**

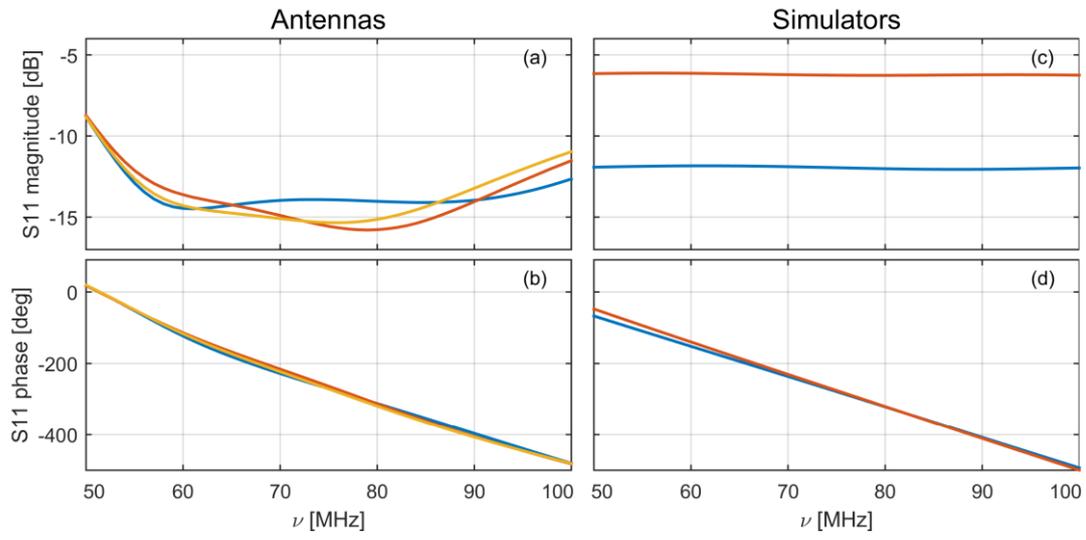

**Extended Data Figure 3**

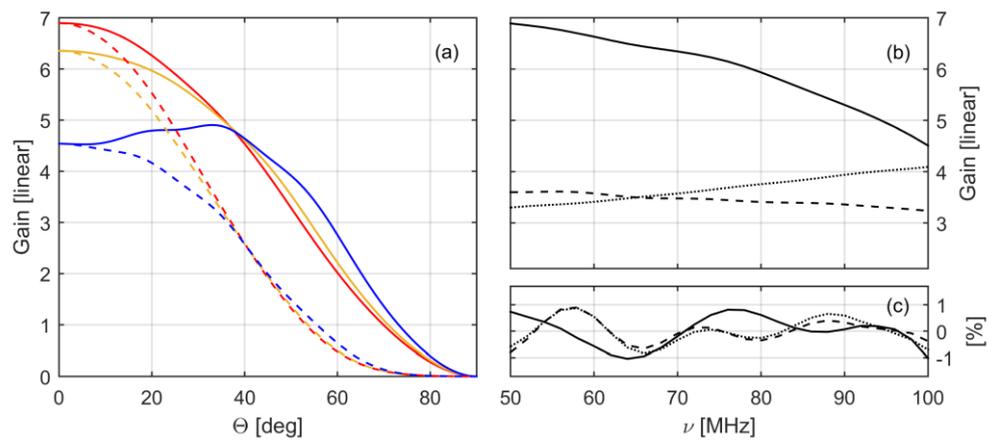

**Extended Data Figure 4**

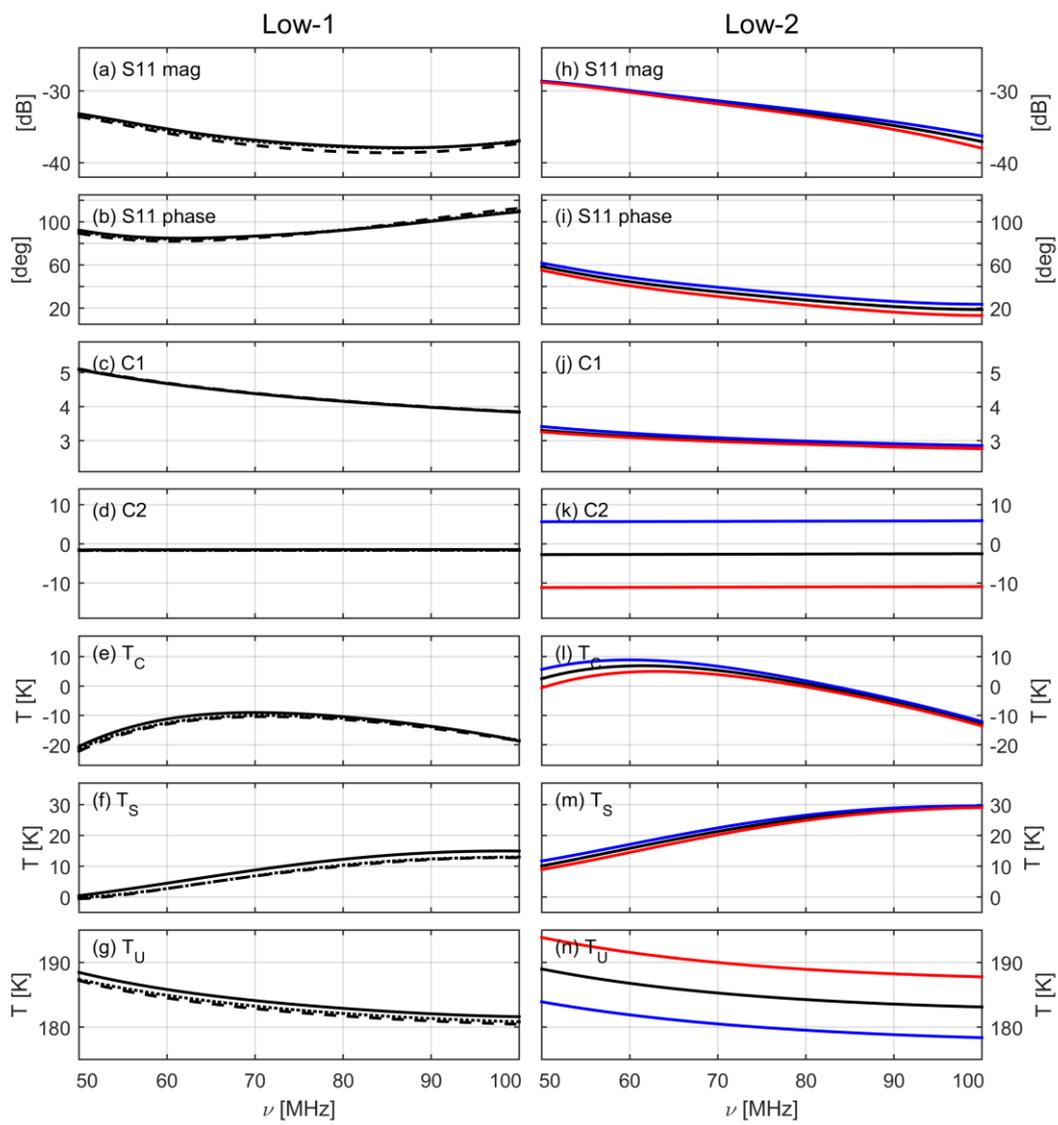

**Extended Data Figure 5**

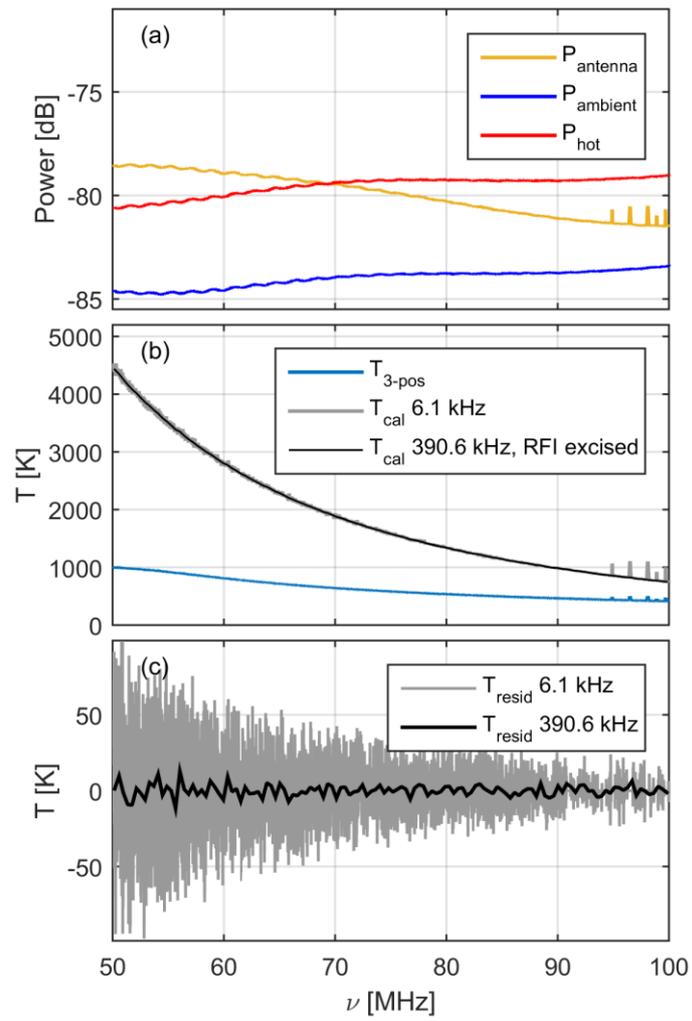

**Extended Data Figure 6**

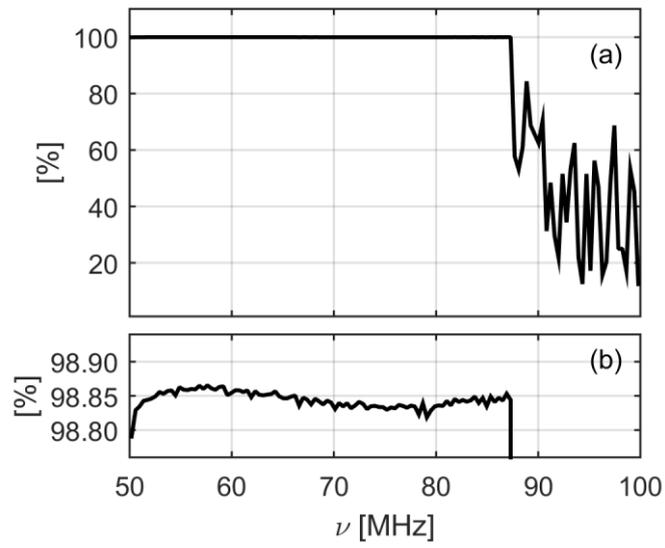

**Extended Data Figure 7**

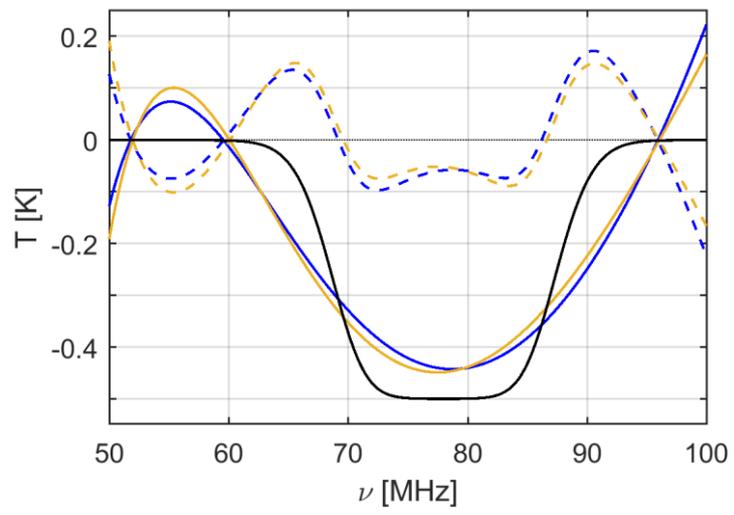

**Extended Data Figure 8**

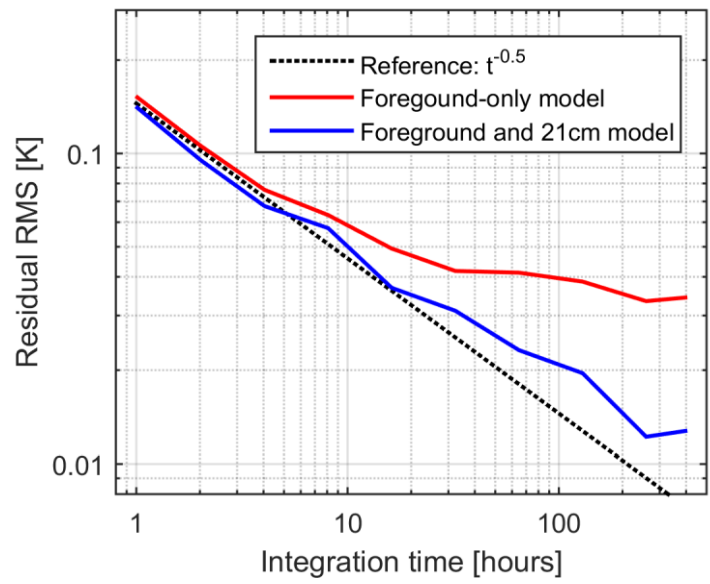

**Extended Data Figure 9**

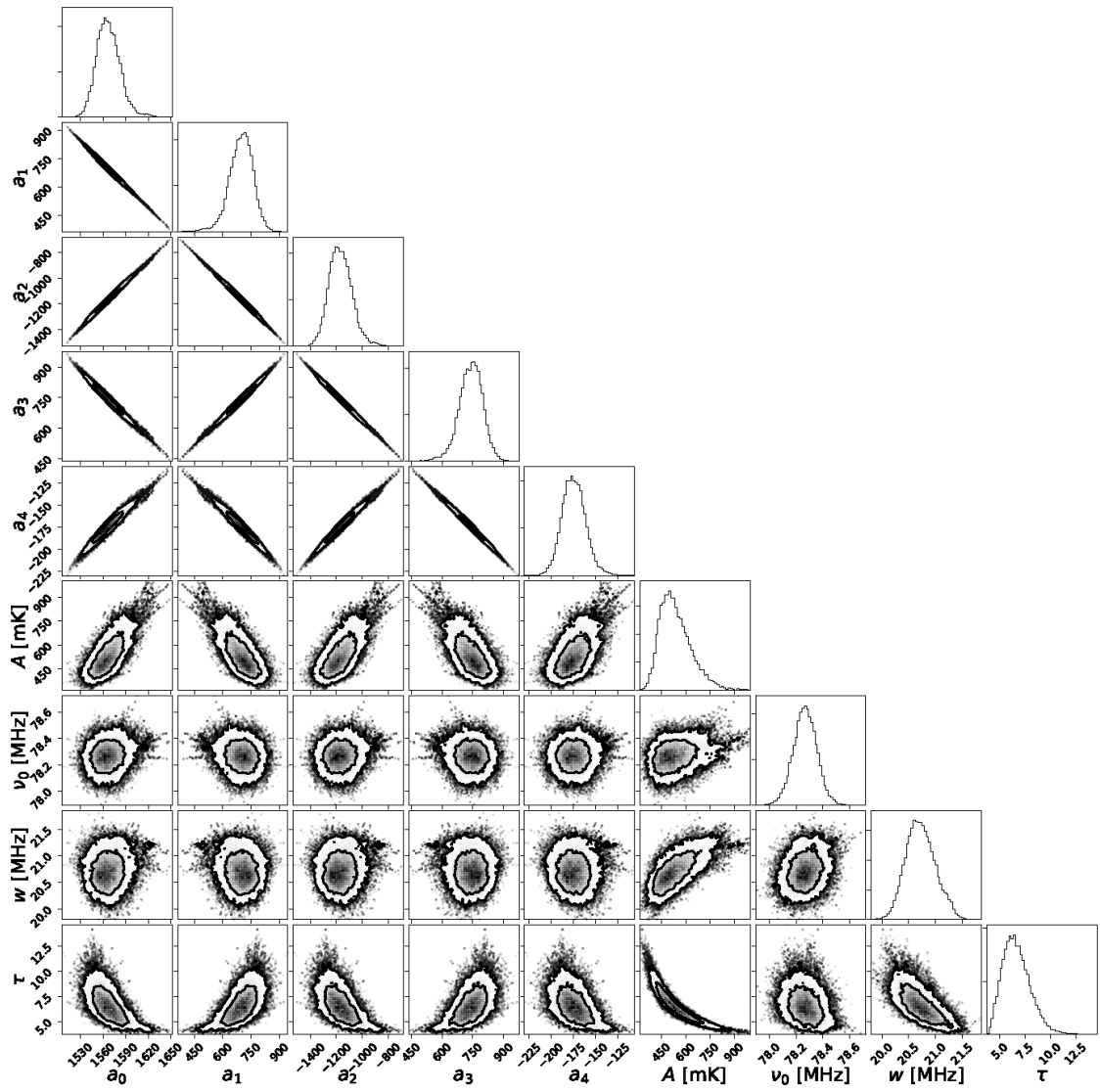

**Extended Data Figure 10**